\pdfoutput=1
\documentclass[a4paper,american,floatfix,pdftex,superscriptaddress,twoside,%
aip,jcp,%
citeautoscript,
reprint,twocolumn
]{revtex4-1}%
\usepackage[T1]{fontenc}
\usepackage{graphicx}%
\usepackage[utf8]{inputenc}
\usepackage{hyperref, hypernat}
\usepackage[todos]{CMImacros}

\graphicspath{{./figures/}} 
\hypersetup{citebordercolor=yellow,linkbordercolor=red,urlbordercolor=blue}

\newcommand{\cfeldesy}{\affiliation{Center for Free-Electron Laser Science, Deutsches
      Elektronen-Synchrotron DESY, Notkestraße~85, 22607 Hamburg, Germany}}%
\newcommand{\uhhcui}{\affiliation{Center for Ultrafast Imaging, Universität Hamburg, Luruper
      Chaussee~149, 22761 Hamburg, Germany}}%
\newcommand{\uhhphys}{\affiliation{Department of Physics, Universität Hamburg, Luruper Chaussee~149,
      22761 Hamburg, Germany}}%
\newcommand{\ruimm}{\altaffiliation[Present address: ]{Radboud University, Institute for Molecules
      and Materials, Heijendaalseweg 135, 6525 AJ Nijmegen, The Netherlands}}%
\newcommand{\jkemail}{\email[Email:~]{jochen.kuepper@cfel.de}}%
\newcommand{\cmiweb}{\homepage[\mbox{website}:~]{https://www.controlled-molecule-imaging.org}}%

\begin{document}
\title{Charge-state distribution of aerosolized nanoparticles}
\author{Jannik~Lübke}\cfeldesy\uhhphys\uhhcui%
\author{Nils~Roth}\cfeldesy\uhhphys%
\author{Lena~Worbs}\cfeldesy\uhhphys%
\author{Daniel A.\ Horke}\ruimm\cfeldesy\uhhcui%
\author{Armando~D.~Estillore}\cfeldesy%
\author{Amit~K.~Samanta}\cfeldesy%
\author{Jochen Küpper}\jkemail\cmiweb\cfeldesy\uhhphys\uhhcui%
\date{\today}%

\begin{abstract}\noindent
   In single particle imaging experiments, beams of individual nanoparticles are exposed to intense
   pulses of x-rays from free-electron lasers to record diffraction patterns of single, isolated
   molecules. The reconstruction for structure determination relies on signal from many identical
   particles. Therefore, well-defined-sample delivery conditions are desired in order to achieve
   sample uniformity, including avoidance of charge polydispersity. We have observed charging of 220
   nm polystyrene particles in an aerosol beam created by a gas-dynamic virtual nozzle focusing
   technique, without intentional charging of the nanoparticles. Here, we present a deflection
   method for detecting and characterizing the charge states of a beam of aerosolized nanoparticles.
   Our analysis of the observed charge-state distribution using optical light-sheet localization
   microscopy and quantitative particle trajectory simulations is consistent with previous
   descriptions of skewed charging probabilities of triboelectrically charged nanoparticles.
\end{abstract}
\maketitle

\section*{Introduction}
Single particle imaging (SPI) experiments utilize x-ray diffractive imaging of individual
nano-objects to determine their structure~\cite{Neutze:Nature406:752, Bogan:NanoLett8:310,
   Barty:ARPC64:415}. On a shot-by-shot basis, a stream of aerosolized sample molecules \eg,
artificial nanoparticles~\cite{Ayyer:Optica8:15} or biological
macromolecules~\cite{Hosseinizadeh:NatMeth14:877}, is delivered into the focus of an FEL x-ray 
beam.
The resulting diffraction patterns of the randomly oriented particles are collected and can
subsequently be reconstructed into an average 3D volume with nanometer
resolution~\cite{Loh:PRE80:026705}. Careful and controlled sample delivery are key to successful SPI
experiments~\cite{Martiel:ActaCrystD75:160}: High-resolution structure reconstruction relies on a
very large number of diffraction patterns of isolated, virtually identical
particles~\cite{Ayyer:Optica7:593}.

Biological macromolecules, like proteins, are intrinsically flexible~\cite{Han:VarNatProt:2019} and
their structural integrity can be decreased by Coulomb repulsion between
charges~\cite{deGraff:Structure24:329}. Therefore, excessive charging of injected biomolecules
during SPI experiments must be avoided to maintain their native structure. Neutralizing soft-x-ray
sources~\cite{Modesto-Lopez:JElectrostat69:357} and careful aerosolization
schemes~\cite{Leney:JASMS28:5, Santambrogio:Proteomics19:1800060} help reduce charges on aerosols
created with electrospray-ionization generators. For another popular atomization method using
gas-dynamic virtual nozzles (GDVN)~\cite{Beyerlein:RSI86:125104}, the gas-focused liquid jets are
assumed to not actively charge the sample and buffer solution~\cite{DePonte:JPD41:195505}.

Triboelectric charging of particles in the gas phase, \ie, after aerosolization, was characterized
elsewhere~\cite{Bierwirth:ChemIngTech93:1}. Although the possibility of triboelectric charging of
nanoparticles in GDVNs was pointed out by some of us before~\cite{Awel:JACR51:133}, for these setups
no detailed understanding of the process nor the extent of charging on individual nanoparticles is
available. Here, we demonstrate a direct method for detecting and characterizing the charge-state
distribution of a beam of aerosolized nanoparticles.

\section*{Experimental methods}
In our experiment, prototypical polystyrene-sphere particles (Alfa Aesar, USA,
$d=\left(220.0\pm17.6\right)\text{~nm}$) were transferred from aqueous suspension with a
concentration of $7.5\cdot10^6$~particles/ml into the gas phase using a GDVN which consisted of a
borosilicate glass capillary (inner diameter 30~\um) fitted within a ceramic micro-injection-molded
ejector tip~\cite{Beyerlein:RSI86:125104}. Liquid-sample-line flow rate of approximately 1~\ulpmin
with helium as sheath gas yielded a hit rate of six to seven particles per camera frame at 20~Hz
frame rate, which allowed single-particle counting; acquisition of one data set took 500~s.

\begin{figure}
   \includegraphics[width=\linewidth]{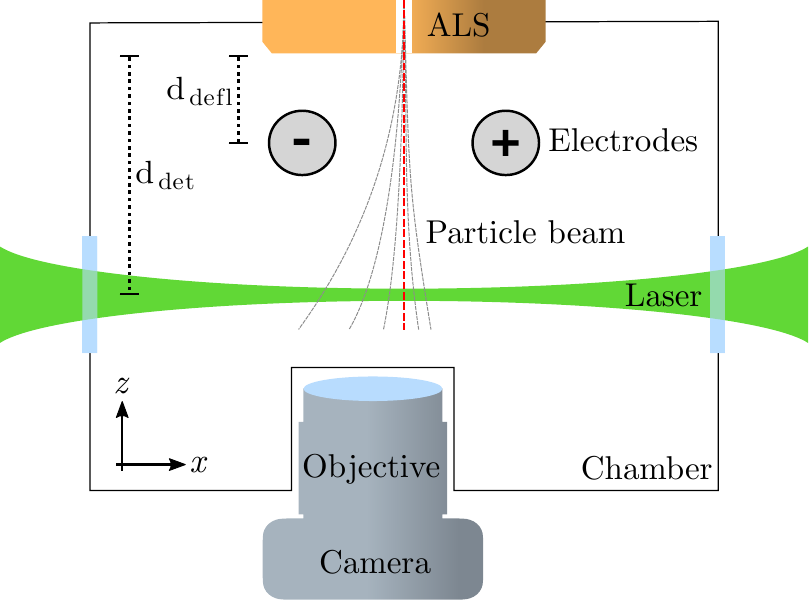}%
   \caption{\textbf{Experimental Setup.} A focused beam of aerosolized nanoparticles (red dashed
      line) is produced using an aerodynamic lens stack (ALS). The particles are deflected (grey
      dashed lines) by the inhomogeneous electric field between two rod-shaped electrodes centered
      in a plane at $d_\text{defl}$ below the ALS. The nanoparticles were detected in a
      position-sensitive light-sheet microscopy setup at $d_\text{det}$ below the ALS. The figure is
      not to scale, see text for details.}
   \label{fig:setup}
\end{figure}

In the gas phase, excess helium was pumped away in a nozzle-skimmer stage~\cite{Roth:JAS124:17} and
the particle beam was focused into the interaction region using an
optimized~\cite{Worbs:geomopt:inprep} aerodynamic-lens stack (ALS)\cite{Liu:AST22:293,
   Roth:flash-nano-injector-polystyrene-beams:inprep}. We used an electrostatic deflector between
the ALS exit and the detection position to disperse the initially cylindrically symmetric
nanoparticle beam according to the particles' charges. The electric field was applied by two 70~mm
long rod-shaped stainless-steel electrodes with a diameter of 4~mm and a center-point distance of
10~mm that were mounted $d_\text{defl}=14.7$~mm below the ALS exit using PEEK holders, see
\autoref{fig:setup}. The ``+'' electrode was connected to positive voltage of a power supply
(Aim-TTi PLH250-P), whereas the ``-'' was connected to floating zero of the voltage supply. This
created an inhomogeneous distorted two-wire field between the two electrodes and the grounded ALS.
The steepest potential gradient between the rod electrodes was $333.3~\text{V}/\text{cm}$.

Individual particles were counted in a size- and position-sensitive light-scattering microscopy
setup~\cite{Awel:OptExp24:6507, Worbs:OptExp27:36580}. The visible-light sheet for detection of the
particles passed through the experimental chamber perpendicular to the injected particle beam
$d_\text{detect}=20$~mm below the last orifice of the ALS. We recorded the scattered light from
intersecting particles using a microscope objective ($5\times$, apochromatic long-working-distance
infinity-corrected objective, Edmund Optics 59-876) and a sensitive camera (Teledyne Photometrics
Prime 95b). The positions of the individual nanoparticles were accumulated into a two-dimensional
(2D) position histogram, yielding a cross section through the particle beam at a given distance from
the injector.
\begin{figure}
   \includegraphics[width=\linewidth]{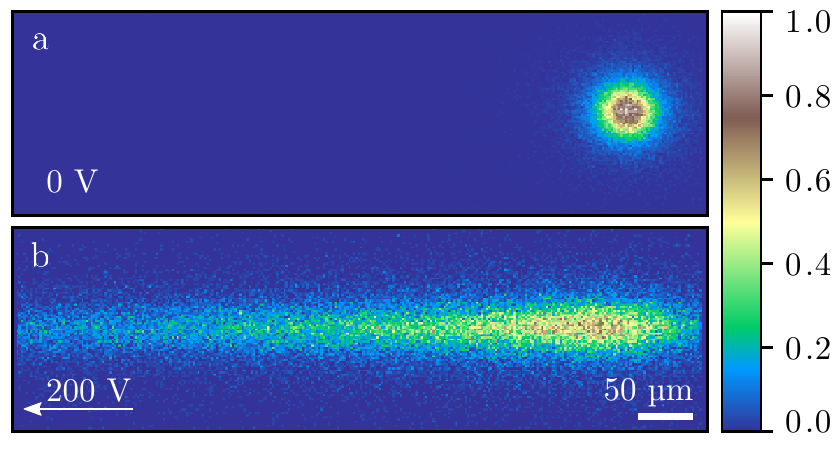}%
   \caption{\textbf{Particle-beam densities.} Particle-beam histograms at $d_\text{detect}=20$~mm
      (a) without and (b) with an electric deflection field of 200~V applied. The white arrow
      indicates the electric-field direction. The colorbar is normalized to the highest counts in
      the 2D histograms.}
   \label{fig:Deflection_profiles}
\end{figure}

\section*{Results and Discussion}
Application of the light-sheet imaging method on the beam of polystyrene nanoparticles yielded the 
particle beam cross sections shown in \autoref{fig:Deflection_profiles}. Without applying an electric 
field, a round particle-beam profile was observed with the highest
density of particles in the center of the beam, \autoref[a]{fig:Deflection_profiles}. When applying
the electric field the beam profile became highly asymmetric along the horizontal field direction,
see \autoref[b]{fig:Deflection_profiles}. The majority of the particles were deflected to the left,
\ie, away from the positive potential, directly implying a charge distribution of significant width
and strongly skewed toward positive charges.
\begin{figure}
   \includegraphics[width=\linewidth]{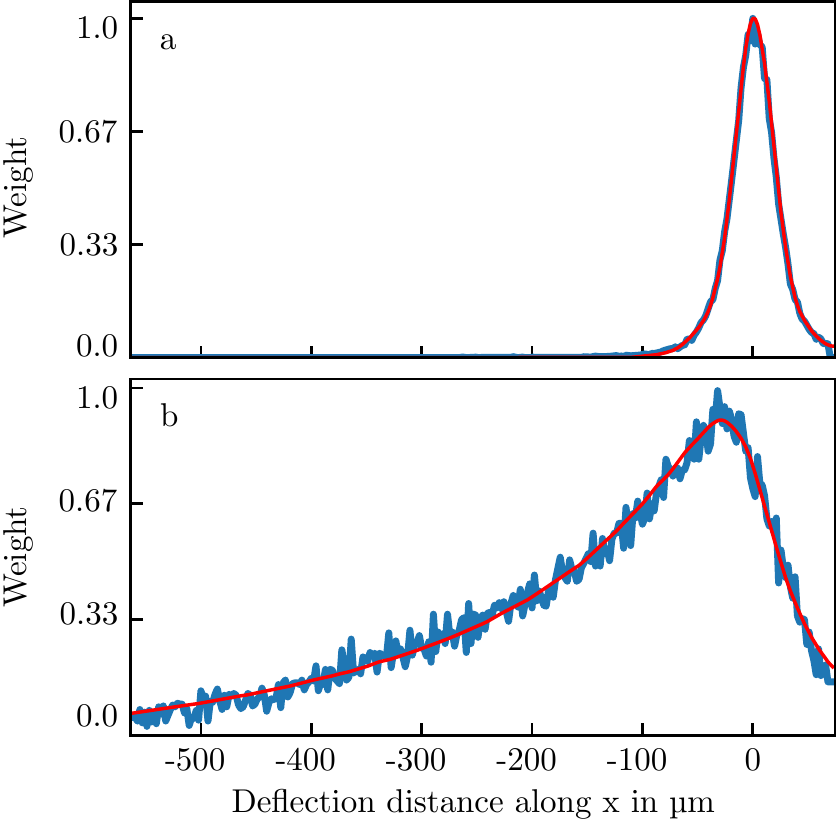}%
   \caption{\textbf{Measured and simulated particle-beam profiles.} One-dimensional particle-beam
      profiles for (a)~0~V and (b)~200~V, \ie, projections of the data in
      \autoref{fig:Deflection_profiles} onto the field axis (blue lines). Corresponding simulated
      particle-beam profiles (red lines) with all profiles normalized to the maximum number of
      events.}
   \label{fig:Meas_and_Sim}
\end{figure}
These density profiles were integrated along the $Y$ direction to yield the experimental beam
profiles shown in \autoref[(blue lines)]{fig:Meas_and_Sim}.

In order to model the observed profiles, we modeled this setup in SIMION~\cite{Simion:8.1} in 2D,
approximating the experimental geometry using two circles for the electrodes and a rectangle for the
injector tip, with the potentials fixed at a given voltage on the ``+'' electrode and 0~V on the
``-'' electrode and the ALS. The 2D approximation of the experimental setup in SIMION was
appropriate, even though the particle beam was cylindrically symmetric compared to translational
symmetry of the electrostatic field: along $Y$, the field can be described as constant, because the
rod electrodes are very long (70 mm) compared to the particle-beam dimensions (31.3~\um at
$d_\text{det}=20$ mm). So, for evaluating the particle deflection only the $Z$ and $X$ coordinates
of the particles need to be considered. For a correct description of the electric far field the
experimental chamber would also needed to be modeled as another ground electrode, but this would
only result in extensive simulation times and can be safely ignored based on our extensive
experience~\cite{Chang:IRPC34:557}.

Using the resulting electric field, we simulated the trajectories of sets of 32000 particles using a
size distribution corresponding to the manufacturer's specifications ($d=(220\pm17.6)$~nm). Based on
particle-beam 3D scans~\cite{Worbs:OptExp27:36580}, we could assume an initial particle-beam
full-width-at-half-maximum (FWHM) of 56.2~\um at the ALS outlet and a width of 29.5~\um in the focus
$18$~mm below the ALS as well as nanoparticle speeds of $v=(130\pm15)$~m/s.

Elementary charges in the range $[-500,500]$ were assigned to each set of particles and the
respective 1001 individual-charge nanoparticle-density profiles were simulated. The simulated
particle positions were sampled at $d_\text{det}=20$~mm and collected into a histogram along $X$,
yielding simulated line profile of the particle beams. We used a bin width of 1.85~\um,
corresponding to the camera pixel edge length in the experimental microscopy setup.

We combined these individual-charge profiles into an overall nanoparticle-density profile as a
weighted average to extract the charge-probability distribution of the nanoparticles. The weights,
\ie, the contributions of the charge states to the sample, were determined from a fit to the
experimental data. We used
\texttt{scipy.optimize.differential\_evolution}~\cite{Virtanen:NatMeth17:261} with a maximum number
of generations of $1000$, a population size of $15$, and a relative tolerance of convergence of
$0.01$ to minimize the mean squared error between the simulated and the measured profiles.

Models for describing the corresponding charge-state distribution included uniform, normal, and
heavy-tailed -- \ie, lognormal, loglogistic, scaled inverse Chi-squared, F, and normal-lognormal
(NLN)~\cite{Yang:ApplEconLett15:737} -- distributions. NLN-distributed charge states were 
described
elsewhere for triboelectrically charged microparticles~\cite{Haeberle:SoftMatter14:4987}. In our
case simulated particle-deflection beam profiles with NLN-distributed charges yielded highest
agreement with experimental profiles. An NLN distribution can be described by the product of two
sets of random variables, one being normally, the other lognormally distributed, which are referred
to by three independent parameters. We fitted these parameters for the retrieval of an NLN
distribution function that reproduced the experimental charge-state distribution very well.

We fitted the charge distribution of the particles in the beam to minimize the $\chi^2$
deviation of experimental and simulated profiles of the particle beam along $X$. Optimization
yielded the simulated profiles in \autoref[(red lines)]{fig:Meas_and_Sim}.We performed the
corresponding $\chi^2$ goodness-of-fit test~\cite{Heckert:NISTstat:2002} with 160 degrees of 
freedom and a pre-determined level of significance $\alpha=0.05$. We calculated $p$-values of 
$p>0.99$. As
$p>\alpha$ there is no significant difference between the measured data and simulations with the
NLN-distributed charge states. Thus the underlying charge-state distribution of the measured
particles can be adequately described by the obtained NLN distribution.
\begin{figure}
   \includegraphics[width=\linewidth]{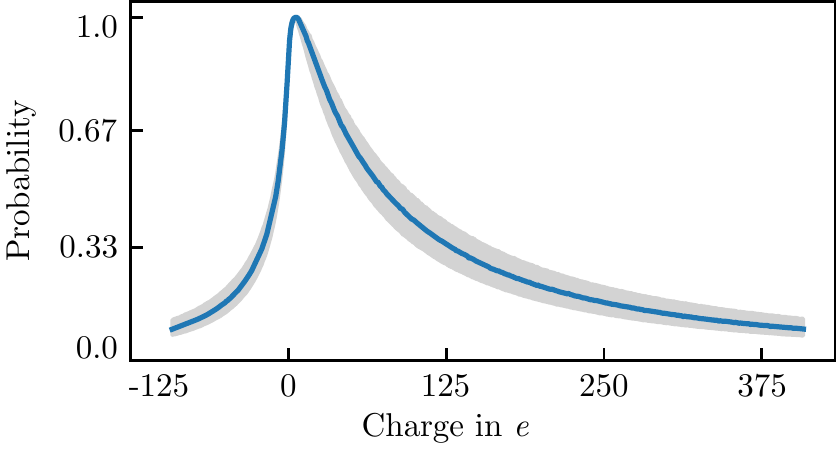}%
   \caption{\textbf{Charge-state distribution.} Histogram of the fitted charge-state distribution in
      the nanoparticle beam, normalized to the counts of the most probable charge of $+5$. The
      $1\sigma$ standard deviation in the probability is shown by the light gray area; see text for
      details.}
   \label{fig:Charge_states}
\end{figure}
The resulting nanoparticle-charge distribution is shown in \autoref{fig:Charge_states}. The most
likely charge observed in our experiment is $+5$. The distribution shows a steep decrease toward
smaller, \ie, negative, and a long tail of positive charges up to $\larger375~e$ on the $220$~nm
polystyrene latex spheres, corresponding to surface charge densities of up to
$+39.5 \text{~nC cm}^{-2}$, which is reasonable (\ie, smaller) compared to published literature
values~\cite{Fernandez:ColloidsSurfA92:121} for positively charged polystyrene nanoparticles.

Positioning uncertainties of the experimental setup and possible deviation of the velocities of the
particles were propagated through the simulations. Here, we assumed uncertainties for the
positioning of the particle beam and the detection light-sheet in respect to the electrodes of
$\pm500$~\um. Furthermore, the mean initial particle-beam velocities were in the range
$\left[120,140\right]$~m/s to incorporate the uncertainty in the simulated velocity of the
particles. These effects result in the uncertainties of the charge probability distribution in
\autoref{fig:Charge_states}

These charges might originate from triboelectric charging in the GDVN~\cite{Awel:JACR51:133}, from
collisions between nanoparticles, or in the aerosol transport tubes due to (multiple) collisions
with the surrounding walls. Even though classical triboelectric charging models would suggest mainly
negative charging of polystyrene on metal surfaces, deviations from the triboelectric series have
been observed~\cite{Shaw:Nature118:659, Shinbrot:epl83:24004, Pham:JElectrostat69:456,
   Xu:ACSNano13:2034}. Another source for particle charging might be the aerosolization process:
during gas focusing of the liquid jet, the collision rates between gas molecules and sample droplets
are high and charge transfer during this process can not be excluded. It is beyond the scope of the
current work to fully determine these physical principles. Instead, we provide a working tool for
future studies when exploring these basics, possibly through systematic experiments about effects of
sample material or aerosolization mechanisms, \eg, electrospray ionization or atomizers, on particle
charging. The current results demonstrate the possibility for controlling and separating charged
particles in SPI experiments and to investigate the effect of defined charges on overall sample
structure and integrity.

\section*{Conclusion}
In conclusion, we have demonstrated a method to characterize the charge-state distribution of a
stream of aerosolized nanoparticles. An ALS injector was used to form a nanoparticle beam. When the
beam was exposed to electrostatic field, we observed large deflection of the nanoparticles,
indicating large charges. We have used charged-particle trajectory simulations to quantitatively
describe our experimental setup. By iteratively fitting the simulated deflection profile with the
experimental one, we have extracted the underlying charge-probability distribution, revealing
significant positive charges ($\larger375~e$). Finding charges on GDVN-aerosolized particles is not
necessarily intuitive. For example during SPI experiments, these particles are presumed to be
overall neutral in charge~\cite{DePonte:JPD41:195505}. Excessive charging can be a source of
structural variance of individual particles, and thus effectively a bottleneck for overall
resolution in structure retrieval. If deemed necessary, neutralizing soft-x-ray devices may be
employed to reduce the overall charges on the (GDVN-) aerosolized nanoparticles. In any case, during
future SPI experiments, it would be highly beneficial to control or select the charge states of the
aerosolized particles, \eg, using the electrostatic deflection technique we presented here.

\section*{Acknowledgments}
This work has been supported by Deutsches Elektronen- Synchrotron (DESY), a member of the Helmholtz
Association (HGF), by the European Research Council under the European Union’s Seventh Framework
Program (FP7/2007-2013) through the Consolidator Grant COMOTION (614507), the Helmholtz Impuls und
Vernetzungsfond, and the Cluster of Excellence ``CUI: Advanced Imaging of Matter'' of the Deutsche
Forschungsgemeinschaft (DFG) (AIM, EXC~2056, ID~390715994).

\bibliography{string,cmi}
\onecolumngrid%
\listofnotes%
\end{document}